\documentclass[10pt,letterpaper]{article}

\usepackage[T1]{fontenc}
\usepackage{lmodern}
\usepackage{microtype}
\usepackage{graphicx}
\usepackage{amsmath,amssymb}
\usepackage{booktabs}
\usepackage{tabularx}
\usepackage{array}
\usepackage{enumitem}
\usepackage[margin=0.85in]{geometry}
\usepackage[hidelinks]{hyperref}

\newcolumntype{Y}{>{\raggedright\arraybackslash}X}

\newenvironment{algorithmblock}[1]
{\par\medskip\noindent\begin{minipage}{\linewidth}\hrule\vspace{0.35em}\textbf{#1}\par\vspace{0.25em}}
{\vspace{0.25em}\hrule\end{minipage}\par\medskip}

\title{\textbf{MeshReduce-U: Compiler-Guided Communication Reduction\\
for Irregular Neural Reductions on Mesh NoCs}}
\author{Amirreza Khorasanian\\
\normalsize Electrical Engineering Student, Department of Electrical and Computer Engineering\\
\normalsize University of Tehran\\
\normalsize \texttt{akh5793@gmail.com}}
\date{Extended preprint}

\begin{document}
\maketitle

\begin{abstract}
Many irregular neural workloads induce skewed many-to-one reductions with repeated neighborhoods and nonlocal communication. Conventional NoC mappers optimize placement and routes for a fixed communication graph, even though associative reductions expose legal opportunities to eliminate traffic before routing. We present MeshReduce-U, a compiler-guided communication-reduction and routing framework for mesh-NoC-based spatial accelerators. MeshReduce-U coalesces colocated sources, forms local aggregation islands, blocks channels with compatible fan-in structure, selects capacity-feasible sinks, and routes the remaining fixed-width carriers using fused usage-aware costs. A deterministic route-replay model reports schedule-derived communication latency, total link usage (TLU), and fused link usage (FusedTLU) separately. Across a 20-workload lowerable neural-network zoo, MeshReduce-U reduces mean latency, TLU, and FusedTLU by 40.3\%, 56.0\%, and 48.7\%, respectively, relative to an ABC-style source-aggregation baseline, improving all three metrics on every workload. Across 40 synthetic irregular reductions, it reduces mean latency and TLU by 12.3\% and 19.7\%. A new 30-instance pass-by-pass study further shows that the structural rewrites reduce the global carrier count by 60.9\% and replay latency by 63.0\%. These results show that rewriting reducible neural communication before routing can be more effective than searching harder over an unreduced traffic graph.
\end{abstract}


\section{Introduction}
Communication is a first-order cost in spatial neural accelerators. Irregular workloads---including graph neural networks (GNNs), sparse feedforward networks, recurrent and gated models, random neural DAGs, and state-space-style models---produce many-to-one reductions with skewed fan-in, repeated neighborhoods, and nonlocal sources. Mapping every contribution as an independent transfer wastes link bandwidth and creates injection and hotspot pressure.

Most NoC mapping methods optimize a fixed communication graph: they move sinks, choose shorter paths, or reroute around congested links. For associative neural reductions, however, the graph itself is not immutable. Contributions originating at one tile can be combined locally; nearby sources can aggregate within a bounded island; and feature channels with the same fan-in pattern can share one block carrier. These transformations remove communication before global routes are chosen.

MeshReduce-U operationalizes this observation as a compiler pass. It first rewrites a lowered many-to-one reduction into fewer physical carriers, then performs capacity-aware sink placement and fused usage-aware routing. We report schedule-derived route-replay latency, ordinary TLU, and FusedTLU separately. This separation matters: route search may lower congestion by spending more links, whereas communication reduction can lower both usage and contention by removing carriers.

Although evaluated on a 2D mesh NoC, the abstraction also matches CGRA-like arrays in which compiler-placed tiles exchange fixed-width payloads over local links. The scope is deliberately layer-local and many-to-one. Whole-DAG residence, one-to-many fanout sharing, and steady-state initiation-interval optimization are outside this formulation.

\textbf{Contributions.} We provide: (1) a formulation that makes legal communication-graph rewriting part of NoC mapping; (2) a conservative stack combining source and island coalescing, feature blocking, usage-aware placement, and fused routing; (3) separate replay latency, TLU, and FusedTLU metrics; (4) a shared-model comparison on 40 synthetic reductions and a 20-workload neural zoo; and (5) a matched pass-by-pass ablation that isolates how the structural rewrites change carrier count and global NoC traffic.

\section{Related Work}
\textbf{NoC mapping and neural gather.} Classical NoC work maps communicating cores to reduce bandwidth cost or congestion while retaining the input communication graph~\cite{dally2001,murali2004}. DNN-specific NoCs have added gather support for abundant many-to-one traffic~\cite{tiwari2022}. MeshReduce-U is complementary: it asks whether the compiler can legally shrink the gather graph before placement and routing.

\textbf{Irregular neural aggregation.} GNN accelerators expose the costs of sparse aggregation, locality, feature reuse, and dynamic sparsity. GNNerator uses feature blocking~\cite{stevens2021}; I-GCN performs runtime islandization~\cite{geng2021}; SCV-GNN improves locality through a sparse representation and ordering~\cite{unnikrishnan2023}; GraphAGILE, DynaSparse, and NeuraChip address broad compiler/runtime, sparsity, and load-balance concerns~\cite{zhang2023graphagile,zhang2023dynasparse,shivdikar2024}. These systems optimize larger accelerator pipelines. MeshReduce-U isolates the layer-local mesh communication problem and combines local aggregation, feature blocking, sink placement, and route selection under one traffic model.

\textbf{Communication reduction and trees.} ABC aggregates before distributed GNN communication~\cite{su2022}, while locality-aware merge methods remove redundant aggregation work~\cite{sun2025}. Steiner-style routing can minimize shared-tree footprint~\cite{kou1981}, but compact trees may contain long or congested paths. We therefore keep ordinary traffic and route-replay latency visible beside fused footprint.

\section{Problem Formulation}
Consider a directed 2D mesh $G=(C,E)$. A lowered layer induces reductions $V$; output $v$ has a multiset of source locations $S_v=\{s_{v,1},\ldots,s_{v,k_v}\}$ and a capacity-feasible sink $c_v\in C$. A branch-explicit lowering routes one path $P_{v,i}$ per source occurrence.

MeshReduce-U distinguishes logical branches $B$ from physical carriers $B'$ produced by legal compiler rewrites, with $|B'|\le |B|$. Ordinary total link usage counts physical-carrier traversals:
\begin{equation}
\mathrm{TLU}(R)=\sum_{b\in B'} |P_b|.
\end{equation}

\begin{figure}[t]
\centering
\includegraphics[width=0.99\linewidth]{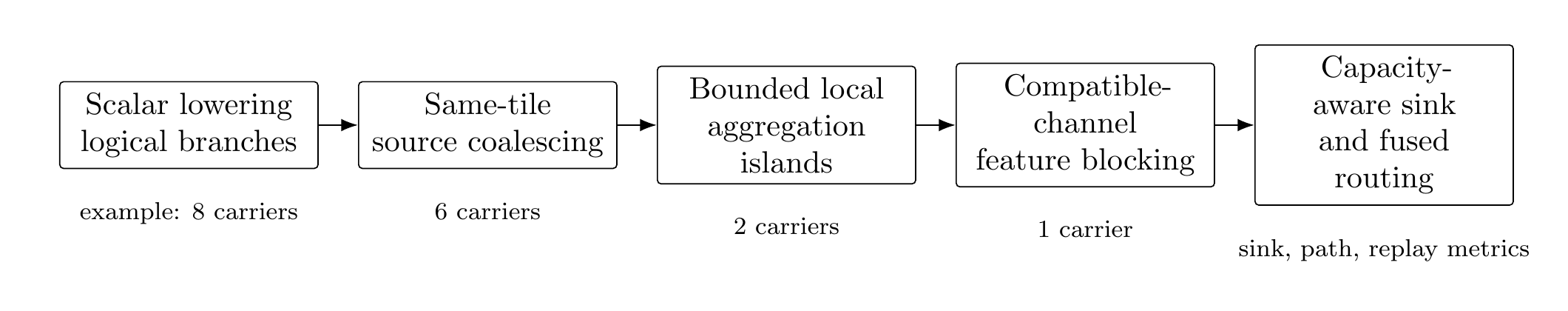}
\caption{MeshReduce-U compresses a legal reduction carrier graph before global placement and routing. The counts illustrate the running example with two compatible channels and four source occurrences per channel.}
\label{fig:pipeline}
\end{figure}

Fused link usage counts a directed edge at most once per output reduction:
\begin{equation}
\mathrm{FusedTLU}(R)=\sum_{e\in E}\left|\left\{v:\exists b\in B'_v,\ e\in P_b\right\}\right|.
\end{equation}
The first metric reflects ordinary routed traffic; the second reflects the footprint if same-output carriers can merge on common tails.

The route-replay model advances each carrier hop by hop on unit-latency directed links. A link serves one carrier per cycle and conflicts serialize in a fixed deterministic order. If $a_b$ is carrier $b$'s final arrival, $T_{\mathrm{sched}}=\max_b a_b$. Model-level latency is the sum over lowered layers. These are abstract communication cycles, not RTL-measured latency; TLU is a carrier-traversal metric rather than an energy claim.

For diagnosis, the implementation records lower bounds from the longest selected path, maximum edge serialization, and maximum source-injection pressure. Their maximum is $T_{\mathrm{LB}}$; the ratio $T_{\mathrm{sched}}/T_{\mathrm{LB}}$ indicates whether unresolved delay is structural or whether more route search may still help. The ratio is a refinement trigger, not a replacement for the three reported metrics.

A carrier denotes one configured fixed-width block payload. Feature channels are packed only when fan-in signatures and payload metadata are compatible and the configured representation can hold the block; otherwise the lowering falls back to smaller or scalar carriers. Thus blocking changes the number of routed transfers, not the mathematical reduction.

\section{MeshReduce-U}
Figure~\ref{fig:pipeline} shows the pipeline. Each transformation is conservative and deterministic for a fixed instance and configuration.

\textbf{Source and island coalescing.} Contributions at the same physical tile and target output are locally reduced. The compiler partitions branches by output identifier, physical source, and compatible payload metadata; exact same-tile duplicates become one carrier. Island coalescing generalizes this step to bounded nearby groups: a local representative is selected and one aggregate exits toward the global sink. The transformation is enabled only when the operator is associative and the frontend marks the intermediate merge as legal. Its purpose is not to approximate a global Steiner tree, but to prevent short, obviously redundant movements from consuming global links.

\textbf{Feature blocking.} Repeated neighborhoods often recur across feature channels. A useful point of precision is that the current implementation does \emph{not} use a geometric ``close enough'' threshold for blocking. After the preceding source/island rewrites, each output channel $v$ is assigned a fan-in signature
\begin{equation}
\sigma(v)=\operatorname{sort}\!\left(S'_v\right),
\end{equation}
where $S'_v$ is the post-coalescing multiset (or set, when duplicates have already been eliminated) of physical source coordinates. In the controlled ablation, outputs are block-eligible only when these signatures are exactly equal. The more general word compatible additionally requires matching operator/payload metadata and that the configured carrier representation can hold the packed channels. Thus locality closeness is an island criterion; feature blocking is a repeated-structure criterion.

\begin{algorithmblock}{Algorithm 1 \quad Compiler-guided reduction and routing}
\textbf{Require:} reductions $V$, source multisets $S_v$, mesh $G$, tile capacities\\
\textbf{Ensure:} capacity-feasible sinks, paths, and replay metrics
\begin{enumerate}[leftmargin=2.2em,label=\arabic*:,nosep]
\item \textbf{for all} $v\in V$ \textbf{do}
\item \quad group operands by physical tile and compatible payload metadata
\item \quad locally reduce each legal same-tile group
\item \quad form bounded legal islands; emit one aggregate per island
\item \textbf{end for}
\item group outputs with compatible fan-in signatures into bounded blocks
\item initialize directed-edge ordinary and fused usage counters
\item \textbf{for all} blocks/reductions in deterministic order \textbf{do}
\item \quad enumerate capacity-feasible sink candidates
\item \quad score each sink by carrier distance, injection pressure, and usage
\item \quad enumerate bounded candidate paths to the selected sink
\item \quad choose the path set minimizing incremental ordinary/fused usage
\item \quad commit paths and update edge counters
\item \textbf{end for}
\item deterministically replay all carriers and report $T_{\mathrm{sched}}$, TLU, FusedTLU
\end{enumerate}
\end{algorithmblock}

This distinction also reveals a genuine optimization trade-off. Packing two channels that share the same signature removes transfer multiplicity, but it also removes the freedom to route those channels along different paths. Under the paper's fixed-width-carrier abstraction, a legal block consumes one service opportunity per traversed link, so blocking normally reduces injected carriers without increasing per-hop service demand. On a physical NoC whose link width is narrower than the packed block, however, a block may require multiple flits and can serialize longer. Even in the abstract model, the remapped block may select a different sink/path and occasionally increase TLU or replay time.

The matched 30-instance ablation makes this non-monotonicity visible. Relative to the preceding source+island stage, feature blocking reduces the mean carrier count from 53.33 to 41.20 and replay from 11.20 to 7.60 cycles, but on a per-instance basis replay improves/ties/worsens in 26/2/2 cases and TLU improves/worsens in 26/4 cases. Hence blocking is highly useful on average but is not a theorem that every downstream metric must improve.

A natural deployment extension is therefore profitability-aware blocking: enumerate a small family of legal partitions of a compatible signature class (for example scalar, pairs, quartets, and the maximum supported block), route/replay each bounded candidate, and select lexicographically by $(T_{\mathrm{sched}},\mathrm{TLU},\mathrm{FusedTLU})$. We do not fold this extension into the frozen headline results; it follows directly from the same candidate-guard philosophy used later for ordinary-versus-fused routing.

\textbf{Placement and routing.} After compression, the mapper evaluates capacity-feasible sinks using aggregate carrier distance and current usage. The selected sink is not necessarily the geometric center: a slightly farther tile can avoid an already pressured cut or injection region. Routing scores candidate paths with both ordinary incremental traversals and fused incremental usage. Common tails are rewarded only when they do not create pathological detours or hot links.

\textbf{Legality and fallback.} Same-tile and island aggregation require a compatible associative operator; feature blocking requires equal or compatible fan-in signatures and payload metadata; sink placement must satisfy output capacity. If a guard fails, that pass leaves the affected branches unchanged. This explicit fallback makes metric changes attributable to the selected policy rather than to run-order noise or approximate semantics.

\begin{table}[t]
\centering
\caption{Legality guards and deterministic fallback behavior.}
\small
\begin{tabularx}{\linewidth}{@{}lYY@{}}
\toprule
Pass & Required guard & Fallback when false\\
\midrule
Source & same output, tile, operator, payload & retain separate carriers\\
Island & legal associative/order-compatible merge & route sources globally\\
Blocking & compatible fan-in signature and width & smaller block or scalar\\
Placement & remaining capacity at selected sink & next feasible sink\\
Routing & bounded legal path exists & deterministic base path\\
\bottomrule
\end{tabularx}
\label{tab:legality}
\end{table}

\begin{table}[t]
\centering
\caption{Role of evaluated MeshReduce variants.}
\small
\begin{tabularx}{\linewidth}{@{}lYY@{}}
\toprule
Variant & Main mechanism & Role in evaluation\\
\midrule
Core & base placement/routing & no full reduction stack\\
U & structural reduction + fused routing & proposed default\\
LU & U + light latency polish & post-reduction search ablation\\
G/L & heavier guarded/aggressive search & synthetic latency frontier\\
\bottomrule
\end{tabularx}
\label{tab:variants}
\end{table}

\textbf{Running example.} One reduction on a $4\times4$ mesh has two feature channels, each with four source occurrences: two at $(0,0)$, one at $(1,0)$, and one at $(1,1)$. Branch-explicit lowering creates eight transfers. Same-tile coalescing removes one duplicate per channel; a local island combines the three nearby tiles; identical fan-in permits the two channels to use one block carrier. The final stage then chooses a capacity-feasible sink and path using current edge usage.

\subsection{Correctness Contract and Cost}
\textbf{Semantic preservation.} For each reduction, the compiler carries an operator identifier, payload type, output identifier, and an explicit merge-legality flag. Same-tile and island passes only re-parent operands within a legal reduction tree; they never delete an operand. Feature blocking only packs independent outputs with compatible signatures into one fixed-width transfer and unpacks them before their separate reductions. Therefore, for an associative operator (and, for order-sensitive associative operators, an order preserved by the frontend), the transformed carrier graph computes the same output values as branch-explicit lowering. Capacity checks alter only the sink choice. If any precondition fails, the corresponding group is left scalar and unchanged.

\textbf{Monotonic structural invariant.} Let $B_i$ be the physical-carrier multiset after structural stage $i$. Source coalescing, island aggregation, and feature blocking construct a partition of the previous operands and emit at most one carrier per legal group, hence $|B_{i+1}|\le |B_i|$. This invariant does not imply that every route is shorter: a usage-aware router may accept a bounded detour to avoid a hot edge. That is why the evaluation reports both carrier count and replay/TLU metrics.

\textbf{Compiler cost.} With $n=|B|$ logical branches, $m=|C|$ candidate sink tiles, $K$ bounded route templates, and maximum candidate length $D$, hash/sort grouping is expected $O(n\log n)$, sink evaluation is $O(|V|m)$, and route scoring is $O(|B'|KD)$. The committed route representation uses $O(|B'|D+|E|)$ space. The heavier G/L variants increase $K$ or repeat search; U keeps these bounds small and is therefore the default stack.

\section{The Algorithmic Core Behind Placement}
The conference-length description calls Stage I ``capacity-aware sink placement.'' The implementation is more algorithmically structured: it combines a bottleneck-radius objective, binary search, exact bipartite max-flow feasibility, and a min-cost-flow tie-break. An alternative total-hop placement is also available, and the default auto policy evaluates both before routing.

\begin{figure}[t]
\centering
\includegraphics[width=0.48\linewidth]{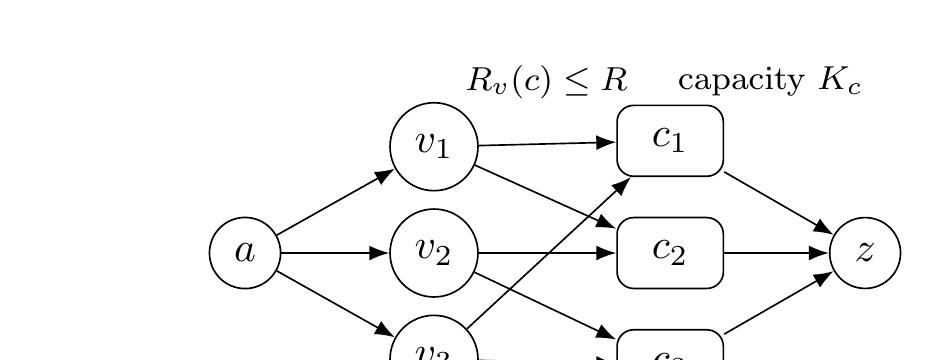}
\caption{Stage-I radius feasibility is a bipartite max-flow problem. Output-to-core arcs exist only for radius-feasible sinks.}
\label{fig:flow}
\end{figure}

\subsection{Min-Max Radius as a Monotone Feasibility Problem}
For output $v$ and candidate sink $c$, define
\begin{equation}
R_v(c)=\max_{s\in S'_v} d(s,c),
\end{equation}
where $d$ is Manhattan distance on the mesh (or shortest-path distance on a general graph). For a radius threshold $R$, output $v$ may use only candidates
\begin{equation}
C_v(R)=\{c\in C:R_v(c)\le R\}.
\end{equation}
The question ``can all outputs be assigned to sinks within radius $R$ without violating tile capacities?'' is an exact flow-feasibility problem, naturally expressed with standard network-flow machinery~\cite{ahuja1993}.

Construct a directed bipartite network with source node $a$, one node for each output $v$, one node for each candidate core $c$, and terminal $z$. Add
\[
a\to v\quad(\mathrm{cap}=1),\qquad v\to c\quad(\mathrm{cap}=1)\ \text{iff }c\in C_v(R),
\]
and
\[
c\to z\quad(\mathrm{cap}=K_c),
\]
where $K_c$ is sink capacity. Radius $R$ is feasible exactly when the maximum $a$--$z$ flow has value $|V|$.

Feasibility is monotone: if $R$ is feasible, every $R'>R$ is feasible because it can only add output-to-core arcs. Therefore a binary search over the integer radius range finds the minimum feasible radius $R^*$ exactly (subject to the candidate set retained by the compiler).

\subsection{Why a Min-Cost Flow Tie-Break?}
Many assignments may achieve the same optimal bottleneck radius $R^*$. The implementation therefore assigns costs to $v\to c$ arcs and solves a min-cost maximum flow on the final feasible graph. Depending on configuration, the tie-break can prefer total source-to-sink hop count, radius, a lexicographic radius-then-hop surrogate, or a bounding-box cost. This cleanly separates two goals: first minimize worst source-to-sink reach, then improve aggregate communication cost.

\begin{algorithmblock}{Algorithm 2 \quad Capacity-feasible minimum-radius placement}
\textbf{Require:} post-rewrite sources $S'_v$, candidate cores $C$, capacities $K_c$
\begin{enumerate}[leftmargin=2.2em,label=\arabic*:,nosep]
\item precompute $R_v(c)$ and optional total distances $H_v(c)=\sum_{s\in S'_v}d(s,c)$
\item $R_{\mathrm{lo}}\leftarrow$ geometric lower bound; $R_{\mathrm{hi}}\leftarrow\max_{v,c}R_v(c)$
\item \textbf{while} $R_{\mathrm{lo}}\le R_{\mathrm{hi}}$ \textbf{do}
\item \quad $R\leftarrow\lfloor(R_{\mathrm{lo}}+R_{\mathrm{hi}})/2\rfloor$
\item \quad build the radius-feasible bipartite flow network
\item \quad \textbf{if} maximum flow has value $|V|$ \textbf{then}
\item \qquad remember $R$; $R_{\mathrm{hi}}\leftarrow R-1$
\item \quad \textbf{else}
\item \qquad $R_{\mathrm{lo}}\leftarrow R+1$
\item \quad \textbf{end if}
\item \textbf{end while}
\item on the $R^*$-feasible graph, run a min-cost max-flow tie-break
\item \textbf{return} the core carrying one unit of flow from each output node
\end{enumerate}
\end{algorithmblock}

The alternative hop-count placement skips radius search and directly solves a capacity-constrained min-cost flow with arc cost
\[
H_v(c)=\sum_{s\in S'_v}d(s,c).
\]
The default auto policy runs both implemented objectives and selects using a cheap congestion/latency proxy before the real routing stage. This is why the high-level phrase ``usage-aware placement'' in the short paper hides both max-flow and min-cost-flow machinery.

\section{The Algorithmic Core Behind Routing}
After placement, the remaining problem is not simply shortest-path routing. We want a bounded family of plausible paths and then a joint assignment that controls the peak directed-edge load. The implementation uses Dijkstra-generated route templates, a thresholded min-max assignment, binary search on the allowed peak load, and optional local hot-edge repair.

\subsection{Dijkstra Trees as Reusable Route Templates}
For a template index $t$, each directed edge $e$ receives a positive weight
\begin{equation}
w_t(e)=\ell(e)\left(1+\epsilon_{t,e}\right)+\alpha p_t(e),
\end{equation}
where $\ell(e)$ is base latency, $\epsilon_{t,e}$ is a bounded deterministic pseudo-random jitter in the U configuration, and $p_t(e)$ is an optional congestion price used by heavier dual-guided variants. For every distinct $(t,\mathrm{sink})$ pair, one reverse Dijkstra run~\cite{dijkstra1959} produces a parent tree oriented toward the sink. Every source assigned to that sink then recovers its path by following parent pointers. Caching the parent tree is important: one shortest-path computation serves all branches ending at that sink.

The jitter is not noise in the evaluation; seeds are fixed. Its purpose is to generate several low-cost but topologically different templates so the assignment stage has route diversity. Heavy variants can instead update $p_t(e)$ from observed congestion and rerun Dijkstra, which resembles a simple dual/pricing method: expensive edges become less attractive in the next template.

\begin{figure}[t]
\centering
\includegraphics[width=0.42\linewidth]{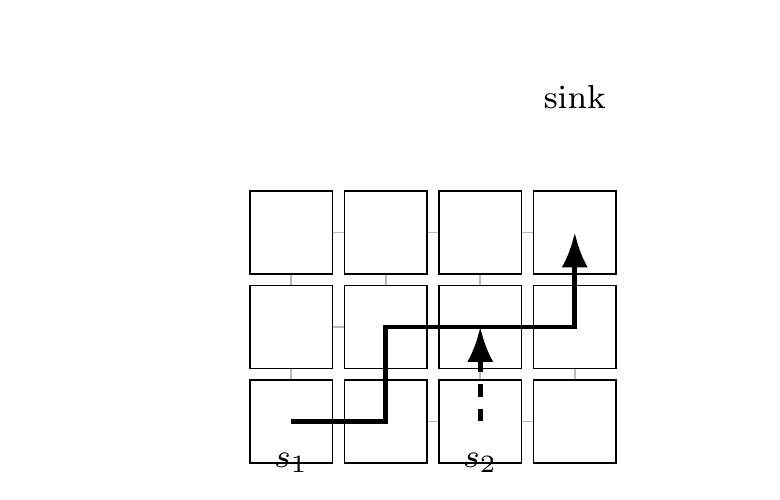}
\caption{One reverse Dijkstra tree can serve many sources of the same sink. Different bounded weight maps generate alternative templates without enumerating arbitrary paths.}
\label{fig:dijkstra}
\end{figure}

\begin{algorithmblock}{Algorithm 3 \quad $K$-template routing with a peak-load threshold}
\textbf{Require:} sinks, $K$ Dijkstra templates, lower/upper thresholds $T_{\mathrm{lb}},T_{\mathrm{ub}}$
\begin{enumerate}[leftmargin=2.2em,label=\arabic*:,nosep]
\item $L\leftarrow T_{\mathrm{lb}}$; $U\leftarrow T_{\mathrm{ub}}$; best $\leftarrow\emptyset$
\item \textbf{while} $L\le U$ \textbf{do}
\item \quad $T\leftarrow\lfloor(L+U)/2\rfloor$
\item \quad greedily assign each carrier to a template without exceeding $T$
\item \quad \textbf{if} all carriers assigned \textbf{then}
\item \qquad best $\leftarrow$ current assignment; $U\leftarrow T-1$
\item \quad \textbf{else}
\item \qquad $L\leftarrow T+1$
\item \quad \textbf{end if}
\item \textbf{end while}
\item optionally reroute users of the hottest edges with bounded-detour Dijkstra
\item \textbf{return} concrete paths and directed-edge loads
\end{enumerate}
\end{algorithmblock}

\subsection{Peak-Load Threshold Search}
Let $P_{b,t}$ be the path of branch/carrier $b$ in template $t$. For a fixed threshold $T$, the assignment oracle processes difficult/low-diversity branches first and chooses a template that keeps every current edge load at most $T$. Among feasible templates, it lexicographically favors the smaller local peak and smaller squared-load increase. In hybrid mode, fused-load terms are added as secondary criteria.

The intended min-max objective is
\begin{equation}
\min_{\tau:B'\to\{1,\ldots,K\}}\ \max_{e\in E}\ \sum_{b\in B'}\mathbf{1}\!\left[e\in P_{b,\tau(b)}\right].
\end{equation}
Rather than optimize this combinatorial assignment exactly, the implementation asks the threshold question ``can the deterministic greedy oracle place all carriers under $T$?'' and binary-searches $T$ between a cut-derived lower bound and a conservative upper bound.

There is an important exactness distinction. Stage I uses an exact max-flow feasibility oracle, so its binary search has the usual monotonic guarantee. Stage II uses a greedy template-assignment oracle: the underlying existence property is monotone in $T$, but the heuristic oracle is not a proof of global feasibility or optimality. We therefore interpret its returned threshold as the best found threshold for the fixed candidate family, not as a certified global optimum. The repository contains separate small-instance certification utilities for checking the gap when desired.

\begin{figure}[t]
\centering
\includegraphics[width=0.58\linewidth]{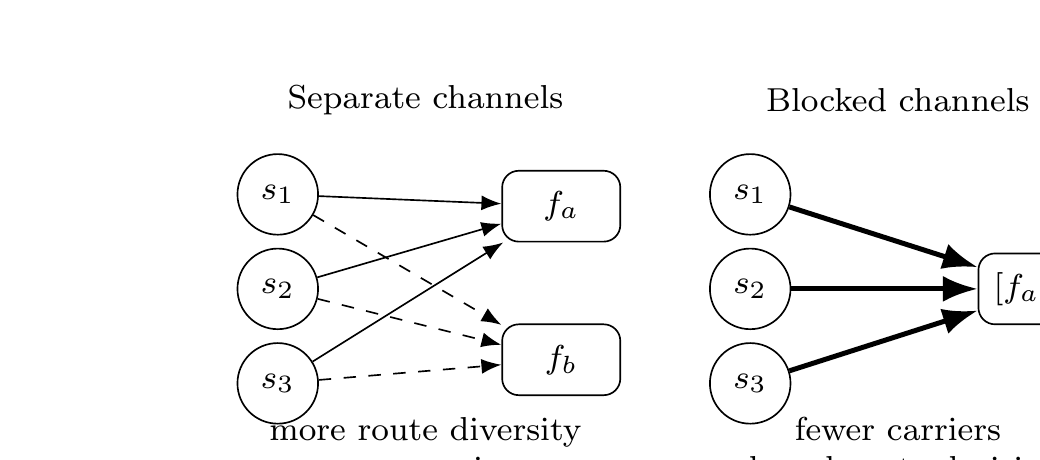}
\caption{Exact-signature feature blocking is legal when metadata/width guards hold, but profitability is a separate routing question.}
\label{fig:blocking}
\end{figure}

\subsection{Hot-Edge Splay and Bounded Detours}
After a candidate assignment, an optional splay pass identifies a maximum-load edge $e^*$, chooses a carrier using it, and reruns Dijkstra while making $e^*$ unavailable (or prohibitively expensive). A move is committed only if the alternative route respects a detour budget and does not create another edge at the current peak load. This is a local-search complement to the global template assignment: the template stage creates diversity cheaply; splay spends extra search only where the realized load reveals a hotspot.

\subsection{Cuts, Lower Bounds, and Search Gating}
The mapper computes cheap lower bounds from longest-path pressure, edge serialization, injection pressure, and mesh cuts; general-graph diagnostics can additionally use a Gomory--Hu cut tree~\cite{gomory1961}. These bounds serve two purposes. First, they show when a mapping is already close to an unavoidable bottleneck. Second, they support the paper's main design rule: expensive route search should be conditional. If replay latency is already near the lower bound after structural reduction, deeper search is unlikely to justify its runtime.

\section{Feature Blocking as a Routing-Diversity Trade-off}
Figure~\ref{fig:blocking} summarizes the distinction between legal blocking and profitable blocking. Two channels with the same fan-in signature can be packed because the same source coordinates are involved, but separate channels can in principle exploit different routes. The current fixed-width model makes packing attractive because a legal block occupies one configured carrier; a flit-accurate implementation should additionally account for payload serialization.

A simple physical extension is to associate a service multiplier
\begin{equation}
\phi(G)=\left\lceil\frac{\sum_{f\in G}w_f}{W_{\mathrm{link}}}\right\rceil
\end{equation}
with a block $G$, where $w_f$ is channel width and $W_{\mathrm{link}}$ is payload capacity per service opportunity. The present experiments effectively restrict legal blocks to $\phi(G)=1$. If $\phi(G)>1$, blocking should be evaluated jointly with routing rather than assumed beneficial. This formulation turns the intuitive question---``could separate channels take better paths?''---into a concrete compiler decision: blocking trades carrier multiplicity for route diversity and possible serialization.

\section{Algorithmic Toolbox and Where Each Piece Matters}
\begin{table}[t]
\centering
\caption{Classical algorithms inside the MeshReduce search stack.}
\scriptsize
\begin{tabularx}{\linewidth}{@{}p{0.18\linewidth}YY@{}}
\toprule
Idea & Role & Why it fits\\
\midrule
Binary search & minimum feasible Stage-I radius; Stage-II load threshold & both are thresholded bottleneck problems\\
Maximum flow & exact capacity feasibility for a radius & outputs need one sink; cores have capacities\\
Min-cost max-flow & tie-break placement / total-hop placement & chooses a globally capacity-feasible low-cost assignment\\
Dijkstra & cached route trees, congestion-priced templates, hot-edge alternatives & positive edge costs and many sources sharing a sink\\
Cut lower bounds & estimate unavoidable congestion & separates structural bottlenecks from search deficiency\\
Gomory--Hu tree & optional general-graph cut diagnostic & summarizes many pairwise min-cuts compactly\\
Steiner/Hanan bias & optional shared-tail candidate generation & exposes merge-friendly geometry without making fused footprint the only metric\\
Greedy + local repair & assign templates and relieve hot edges & keeps the default mapper fast after graph reduction\\
\bottomrule
\end{tabularx}
\label{tab:toolbox}
\end{table}

Table~\ref{tab:toolbox} makes explicit the classical algorithmic ideas that are compressed into a few sentences in the conference paper. They are not decorative implementation details: each solves a different combinatorial subproblem.

The unifying pattern is reduce, bound, then search. Compiler rewrites first shrink the number of carriers. Flow algorithms place the surviving reductions under hard capacity constraints. Shortest-path algorithms generate structured routing choices. Bounds tell us when further search is worthwhile. This composition is the algorithmic reason U can remain fast while heavier G/L variants explore a larger route space.

\section{Experimental Method}
\textbf{Common model.} Every method receives the same serialized instance, directed mesh, source locations, output capacities, and deterministic route-replay model. Links have unit hop latency and one carrier service opportunity per cycle. Instance metadata records dimensions, capacities, and generation seeds. FusedTLU additionally assumes reduction-compatible tags and local accumulation at legal merge points, so ordinary TLU is always reported beside it.

\textbf{Workloads.} The frozen heavy suite contains 40 synthetic/GNN-like reductions stressing skewed fan-in, hotspots, nonlocal sources, and graph-like neighborhoods. The 20-workload zoo lowers GCN, GraphSAGE, GIN, SGC, APPNP, sparse and random DAGs, feedforward and autoencoder representatives, recurrent/gated and reservoir proxies, diffusion-style denoisers, neural-field MLPs, spiking surrogates, state-space proxies, transformer-mixer proxies, and CNN controls. These are compiler-lowerable communication representatives, not claims of full end-to-end model execution.

\textbf{Baselines.} We implement row-major XY, unicast, hop-count placement, load-aware routing, ABC-style gather, I-GCN-style islandization, min-cut locality, and a NetworkX Steiner approximation. They are objective-family proxies rather than full external accelerator ports. All operate on the same lowered traffic, topology, and metrics, isolating communication construction and route policy from unrelated memory, precision, frequency, and platform choices.

\begin{table}[t]
\centering
\caption{Evaluated groups and baseline boundary.}
\small
\begin{tabularx}{\linewidth}{@{}l c Y@{}}
\toprule
Group & Cases & Stress or objective represented\\
\midrule
Synthetic/GNN-like & 40 & skew, hotspots, nonlocal fan-in\\
GNN-style subset & 5 & repeated neighborhoods/features\\
Broad lowerable zoo & 20 & sparse DAGs, gated/mixer/CNN controls\\
Row-major/unicast & -- & fixed routes/no reduction sharing\\
A3MAP/NMAP & -- & distance/load-aware mapping\\
ABC/I-GCN & -- & source aggregation/island locality\\
Partition/Steiner & -- & graph locality/fused footprint\\
\bottomrule
\end{tabularx}
\label{tab:groups}
\end{table}

\textbf{Controlled pass ablation.} To replace a proxy-only mechanism argument, we additionally generate 30 matched associative reductions on a $4\times4$ mesh: 24 outputs, power-law fan-in in $[3,10]$, three source distributions (hotspot, power-law, uniform), and seeds 0--9. The same instances are progressively rewritten by exact source deduplication, radius-1 island aggregation, signature blocking of up to eight outputs, and finally fused/hybrid usage-aware routing. The study isolates the global carrier graph after legal local merges; all stages use the same deterministic replay.

\textbf{Reproducibility.} The implementation separates lowering, communication rewriting, placement/routing, and replay. The same serialized input is supplied to every baseline, and frozen aggregate CSVs are generated from one designated evidence tier. Smoke tests exercise route export and cycle-level replay without changing the paper tables.

\subsection{Measurement Protocol}
Each reported row is produced from a complete mapping followed by route replay; search-time surrogates are not substituted for headline latency. Paths, selected sinks, and directed-edge loads are exported before aggregation, and the replay implementation is separately checked on tiny hand-inspectable instances. All tie breaks, instance seeds, and route orders are fixed. We use arithmetic means for aggregate tables and also report per-instance win counts where available. No method is allowed to change the mesh, output capacities, lowered operands, or replay scheduler.

The comparison intentionally has a narrow validity target: it asks how communication-graph construction and routing objectives behave under one common mesh model. It does not equate proxy rows with complete published accelerators or compare their end-to-end throughput. Conversely, ordinary TLU is retained beside FusedTLU so that U cannot claim a gain solely from idealized in-network merging.

\section{Results}
\subsection{Synthetic Irregular Reductions}
Table~\ref{tab:synthetic} summarizes the 40 cases. MeshReduce-G and MeshReduce-L define the latency-oriented frontier, whereas MeshReduce-U gives the lowest practical TLU. Relative to ABC, U lowers mean replay latency from 31.63 to 27.75 (12.3\%) and TLU from 656.83 to 527.43 (19.7\%); it improves TLU in 40/40 cases and latency in 35/40. Steiner obtains the smallest fused footprint but substantially worsens both latency and ordinary TLU, demonstrating why fused usage cannot stand alone.

\begin{table}[t]
\centering
\caption{Synthetic/GNN-like aggregate; lower is better. Runtime is mapper time in seconds.}
\small
\begin{tabular}{@{}lrrrr@{}}
\toprule
Method & Lat. & TLU & Fused & Time\\
\midrule
MeshReduce-G & \textbf{22.95} & 662.28 & 456.63 & 30.99\\
MeshReduce-L & 23.00 & 682.58 & 468.38 & 19.15\\
MeshReduce-Core & 23.80 & 671.73 & 460.55 & 0.84\\
MeshReduce-LU & 27.35 & 540.38 & 425.32 & 0.54\\
MeshReduce-U & 27.75 & \textbf{527.43} & 393.38 & 0.51\\
ABC fused gather & 31.63 & 656.83 & 424.55 & 0.14\\
I-GCN proxy & 35.43 & 606.58 & 458.98 & \textbf{0.03}\\
NetworkX Steiner & 64.73 & 1043.18 & \textbf{349.55} & 0.20\\
\bottomrule
\end{tabular}
\label{tab:synthetic}
\end{table}

\begin{table}[t]
\centering
\caption{Broad lowerable neural-workload aggregate; lower is better.}
\small
\begin{tabular}{@{}lrrr@{}}
\toprule
Method & Lat. & TLU & Fused\\
\midrule
MeshReduce-U & \textbf{62.70} & \textbf{884.80} & \textbf{682.50}\\
MeshReduce-LU & 63.65 & 897.10 & 696.65\\
MeshReduce-G & 93.00 & 2047.25 & 1403.15\\
ABC fused gather & 105.05 & 2011.65 & 1329.50\\
MeshReduce-L & 247.80 & 6791.15 & 1684.65\\
NMAP proxy & 264.70 & 9678.40 & 2968.90\\
MeshReduce-Core & 276.50 & 6598.35 & 1516.50\\
NetworkX Steiner & 404.55 & 7381.20 & 1216.80\\
\bottomrule
\end{tabular}
\label{tab:broad}
\end{table}

\subsection{Lowerable Neural-Workload Zoo}
On the 20-workload zoo, U is best among evaluated practical methods on all three aggregate metrics (Table~\ref{tab:broad}). Relative to ABC, it lowers mean latency by 40.3\%, TLU by 56.0\%, and FusedTLU by 48.7\%; all three improve for 20/20 workloads. Ordinary TLU also improves strongly, so the conclusion does not depend on idealized in-network fusion.

\subsection{Pass-by-Pass Structural Evidence}
\begin{figure}[t]
\centering
\includegraphics[width=0.92\linewidth]{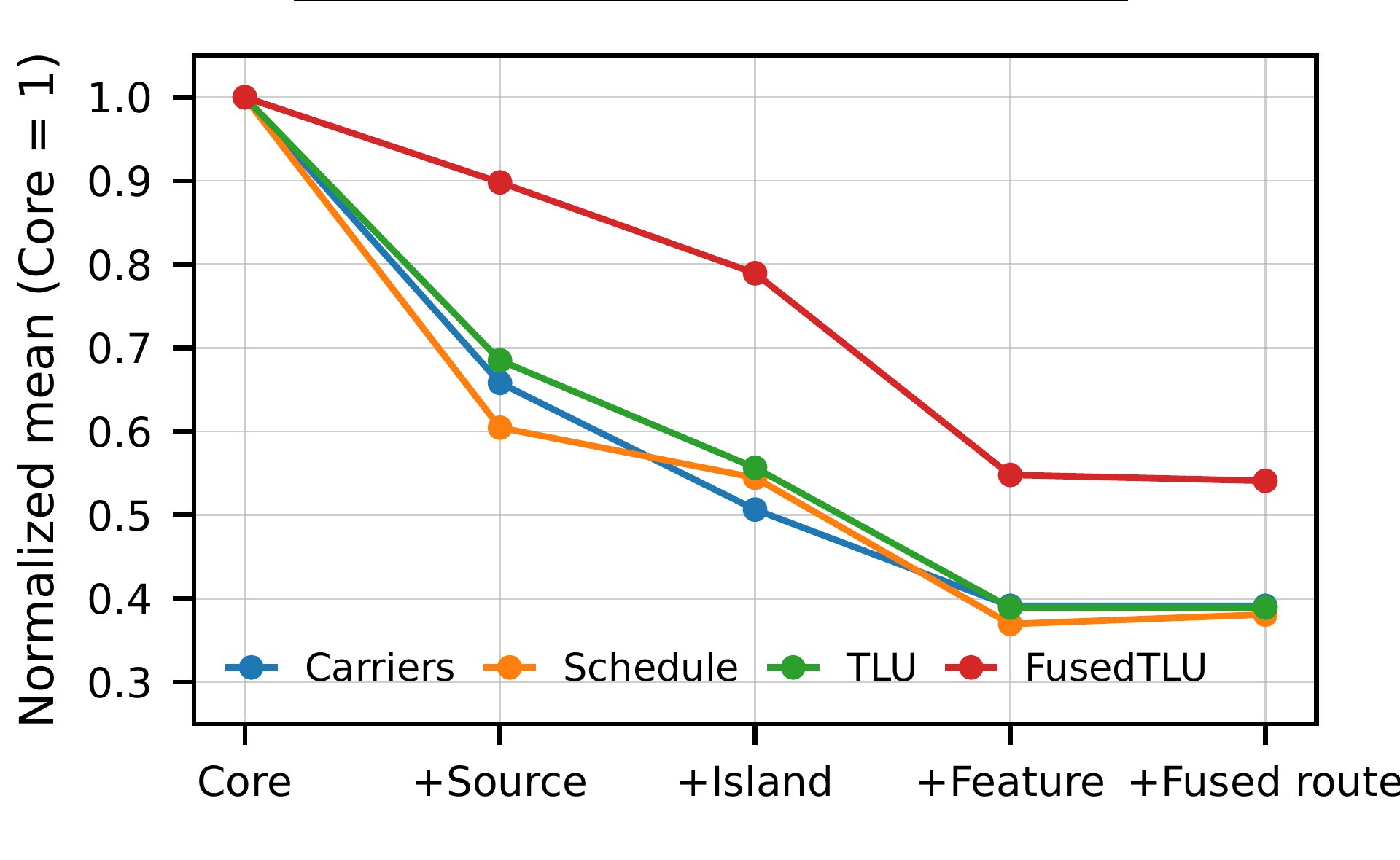}
\caption{Normalized pass-by-pass means for the 30 matched reductions. Structural rewriting, not deeper route search, accounts for the dominant reduction in carriers, replay cycles, and link usage.}
\label{fig:ablation}
\end{figure}

Table~\ref{tab:ablation} and Fig.~\ref{fig:ablation} apply the U passes progressively to the 30 matched reductions. Exact same-tile coalescing removes 34.2\% of carriers on average. Adding islands removes 49.4\% relative to Core. Feature/signature blocking reaches a 60.9\% carrier reduction and lowers mean replay latency from 20.57 to 7.60 cycles (63.0\%) and TLU from 189.03 to 73.57 (61.1\%). This pre-routing structural stage improves replay latency in 30/30 cases.

The final fused-aware routing step preserves carrier count and TLU, lowers mean FusedTLU from 67.20 to 66.33, and changes replay latency from 7.60 to 7.83 cycles. Thus the compiler rewrites supply most of the gain; fused routing makes a small footprint/latency trade-off rather than hiding it. Full U still improves latency in 29/30 cases and both TLU metrics in 30/30 versus Core.

\begin{table}[t]
\centering
\caption{True structural ablation on 30 matched $4\times4$ reductions. Means are global-carrier metrics after each legal rewrite.}
\small
\begin{tabular}{@{}lrrrr@{}}
\toprule
Stage & Carriers & Sched. & TLU & Fused\\
\midrule
Core & 105.30 & 20.57 & 189.03 & 122.63\\
+ source coalescing & 69.30 & 12.43 & 129.50 & 110.13\\
+ island aggregation & 53.33 & 11.20 & 105.20 & 96.80\\
+ feature blocking & 41.20 & \textbf{7.60} & \textbf{73.57} & 67.20\\
+ fused-aware routing & \textbf{41.20} & 7.83 & \textbf{73.57} & \textbf{66.33}\\
\bottomrule
\end{tabular}
\label{tab:ablation}
\end{table}

\subsection{Sensitivity to Source Locality}
The matched ablation deliberately includes three source distributions. Table~\ref{tab:locality} shows that U is not obtaining its mean result from only one favorable pattern. Hotspot and power-law cases expose the most repeated locality, so U removes roughly three quarters of their carriers and TLU and cuts replay latency by 68--80\%. Uniform sources offer fewer legal coalescing opportunities, yet U still lowers carrier count by 34.4\% and TLU by 30.5\%. The one full-U latency regression among 30 cases occurs in this uniform group and reflects the bounded fused-footprint detour discussed above.

\begin{table}[t]
\centering
\caption{Full-U improvement over Core by source distribution (10 matched seeds each).}
\small
\begin{tabular}{@{}lrrr@{}}
\toprule
Distribution & Carriers & Replay cycles & TLU\\
\midrule
Hotspot & $-72.9\%$ & $-80.2\%$ & $-77.5\%$\\
Power-law & $-74.6\%$ & $-68.4\%$ & $-76.1\%$\\
Uniform & $-34.4\%$ & $-25.9\%$ & $-30.5\%$\\
\bottomrule
\end{tabular}
\label{tab:locality}
\end{table}

\begin{figure}[t]
\centering
\includegraphics[width=0.92\linewidth]{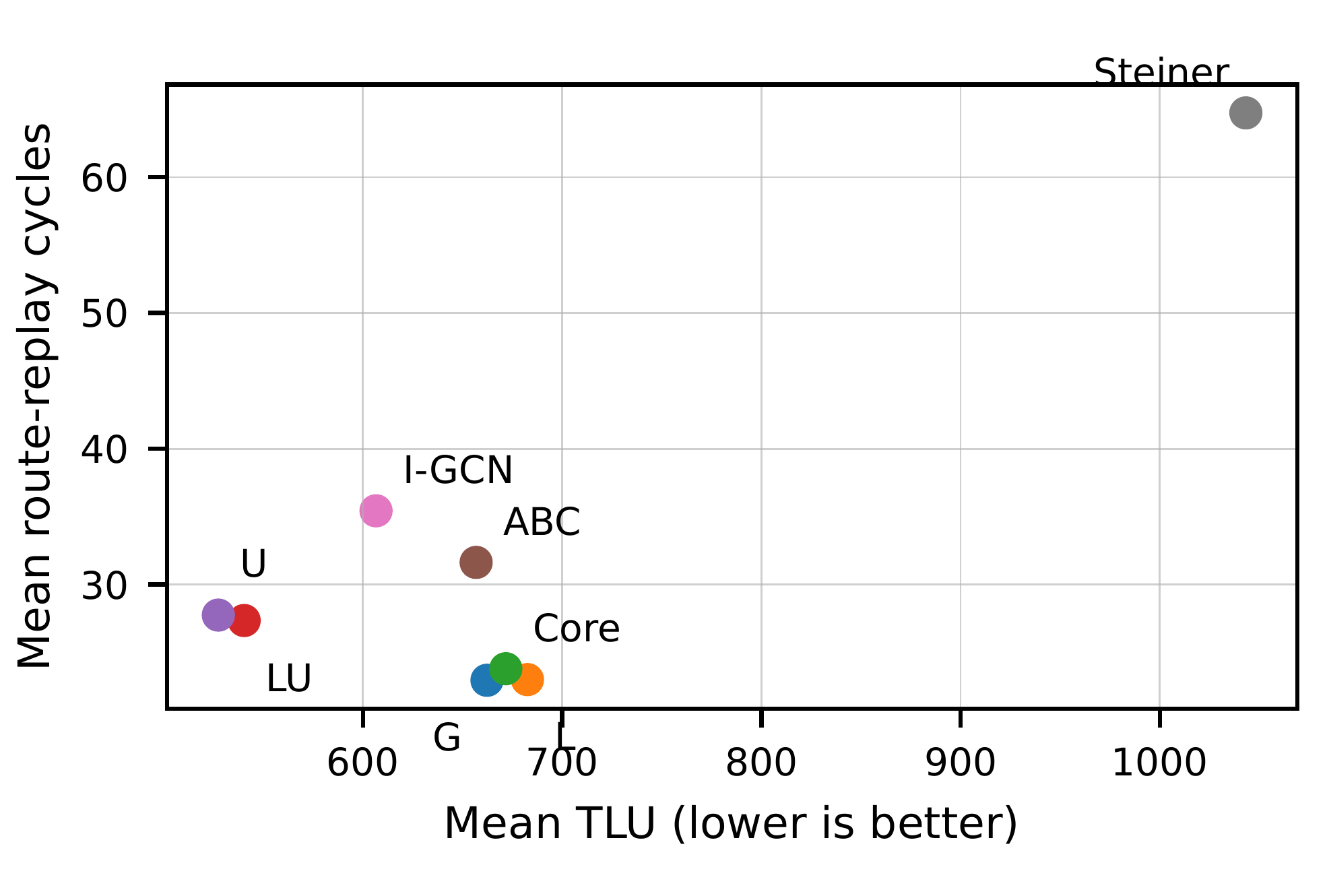}
\caption{Synthetic latency--TLU frontier. G/L spend substantially more search to improve hard-case latency; U minimizes ordinary traffic among practical methods.}
\label{fig:synthetic-frontier}
\end{figure}

\subsection{A No-Regret Candidate Guard}
The single latency regression in the 30-case study is not caused by structural reduction: on uniform seed 3, feature blocking lowers replay from 14 to 11 cycles, while the subsequent fused-aware route changes it to 16 cycles even though TLU falls from 150 to 123. Because both the ordinary-usage and fused-aware path candidates are already available, a compiler can replay both and select
\begin{equation}
R^*=\arg\min_{R\in\{R_{\mathrm{ord}},R_{\mathrm{fused}}\}}\left(T_{\mathrm{sched}}(R),\mathrm{TLU}(R),\mathrm{FusedTLU}(R)\right)
\end{equation}
lexicographically. This guard is deterministic and cannot worsen replay latency relative to either candidate.

Applied post hoc to the matched rows, the guard selects the fused candidate in 8/30 cases and the ordinary candidate in 22/30. Mean replay falls from 7.83 to 7.53 cycles while TLU remains 73.57; FusedTLU rises slightly from 66.33 to 66.60. It improves replay in 30/30 cases versus Core. We keep the frozen main-suite U results unchanged, but this result identifies a low-cost, no-regret deployment refinement rather than motivating universally deeper search.

\begin{table}[t]
\centering
\caption{Candidate guard on the 30-case ablation (means).}
\small
\begin{tabular}{@{}lrrrr@{}}
\toprule
Routing choice & Carriers & Sched. & TLU & Fused\\
\midrule
Ordinary candidate & 41.20 & 7.60 & 73.57 & 67.20\\
Fused candidate & 41.20 & 7.83 & 73.57 & \textbf{66.33}\\
Replay-guarded choice & 41.20 & \textbf{7.53} & 73.57 & 66.60\\
\bottomrule
\end{tabular}
\label{tab:guard}
\end{table}

\begin{figure}[t]
\centering
\includegraphics[width=0.92\linewidth]{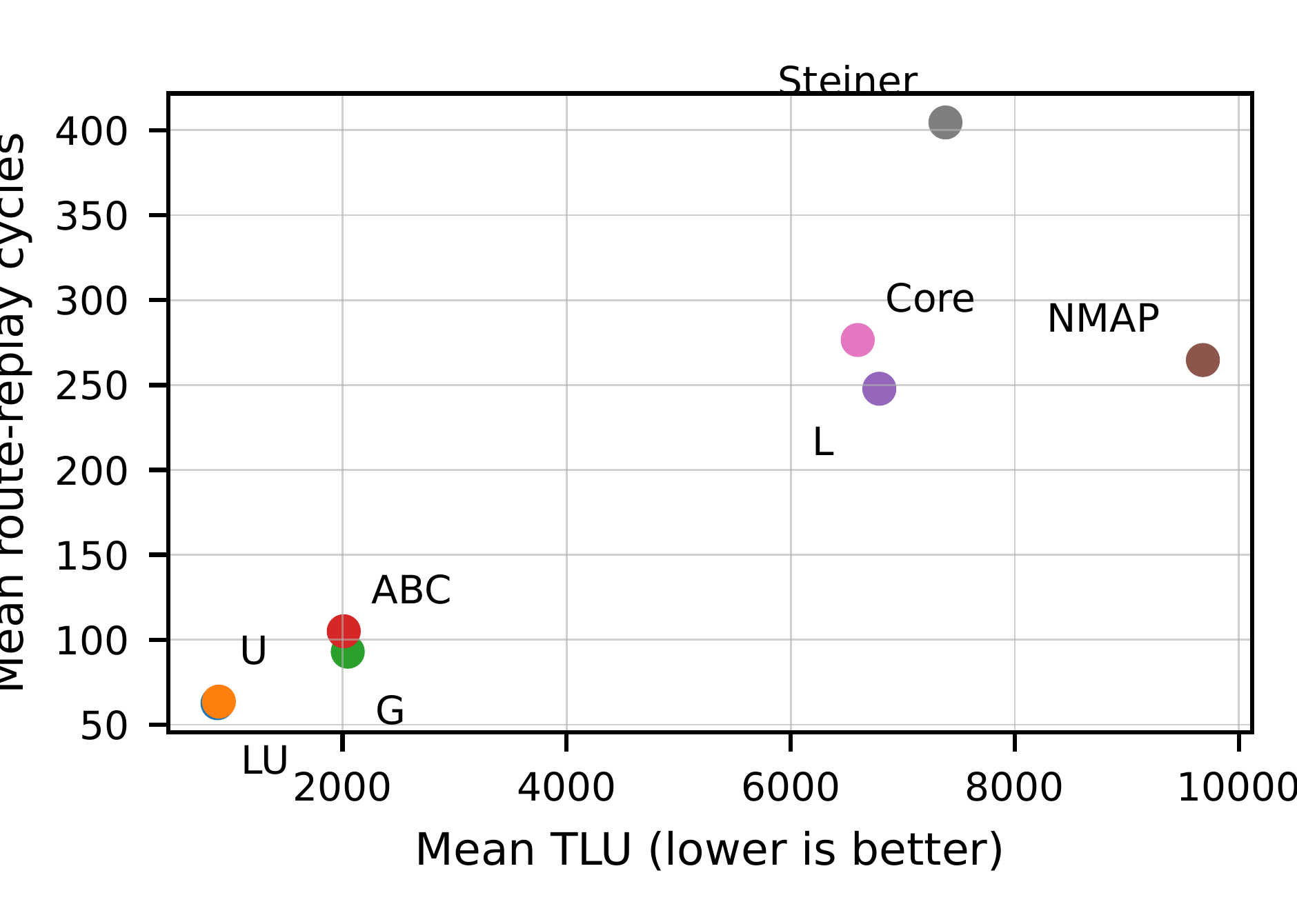}
\caption{Broad-zoo latency--TLU view. U and LU separate sharply from routing-only and fused-footprint alternatives because they remove carriers before global routing.}
\label{fig:broad-frontier}
\end{figure}

\subsection{Route Search after Reduction}
LU applies additional route search after U. On the synthetic suite it improves latency by only 0.40 cycles while increasing TLU by 12.95 and FusedTLU by 31.94. On the broad zoo, LU is slightly worse than U on all three aggregate metrics. Heavy G/L variants help synthetic hard cases, but their 19--31 second mapper times are two orders of magnitude above U. The evidence supports a conditional policy: enable inexpensive reduction by default and trigger heavier search only when lower-bound gaps or concentrated edge pressure reveal unresolved congestion.

\section{Compiler Integration}
The pass consumes a layer-local reduction IR rather than a model-specific graph format. Each operand record contains an output identifier, physical source, payload width/type, reduction operator, and merge-legality metadata; each output exposes feasible sink capacity. The pass returns the rewritten carrier set, selected sinks, directed paths, and replay diagnostics. This interface lets a larger accelerator compiler place compute and memory independently while invoking MeshReduce only for communication-bearing reductions.

The separation also supports gradual hardware adoption. A substrate with only endpoint accumulation can use same-tile coalescing and feature blocking while disabling islands and fused tails. A design with legal intermediate accumulators can enable all passes. Since every unsupported group falls back to the branch-explicit representation, adding a new operator or payload format does not require weakening correctness for already supported layers.

\section{Hardware Realization Contract}
MeshReduce is a compiler pass, but each transformation has a concrete substrate requirement. Table~\ref{tab:modes} separates what can be deployed on a conventional endpoint-accumulation NoC from what needs intermediate merge support. This distinction is also why the paper never converts TLU into an energy number: the same carrier graph can be realized by different routers, buffers, links, and local accumulators.

\textbf{Carrier and tag interface.} A physical carrier contains a bounded payload block plus output and reduction tags. Blocking packs independent channels; it does not add them together. At an endpoint or legal merge point, the tag selects the destination accumulator and operator. The compiler can therefore preserve scalar semantics while reducing the number of configured transfers. Mixed-width or incompatible payloads simply form separate blocks.

\textbf{Routing interface.} The method does not require online adaptive routing. The compiler emits a sink and a directed path for each surviving carrier, and a static or source-routed NoC can follow those paths. Usage-aware scoring is performed offline. A platform may still use virtual channels or buffering, but those choices are outside the abstract replay model and should be calibrated separately.

\textbf{Local aggregation accounting.} Same-tile coalescing is naturally absorbed by the source PE. Island aggregation is more substrate-dependent: an implementation may use a nearby PE, a switch-local arithmetic unit, or a scratchpad reduction engine. The controlled ablation intentionally measures the resulting global carrier graph after legal local merges. It therefore establishes the potential reduction in global traffic, not a claim that local aggregation is free.

\begin{table}[t]
\centering
\caption{Incremental deployment modes. Ordinary TLU remains valid in every mode.}
\small
\begin{tabularx}{\linewidth}{@{}lY@{}}
\toprule
Mode & Enabled behavior and interpretation\\
\midrule
Endpoint-only & source coalescing, blocking, placement, and ordinary routing; FusedTLU is diagnostic\\
Island merge & adds bounded island aggregation; the integrator must account for local merge cost\\
Merge-capable tail & adds common-tail-aware routing; FusedTLU estimates configured shared footprint\\
Replay guard & compiles two bounded candidates and selects without worsening modeled latency\\
\bottomrule
\end{tabularx}
\label{tab:modes}
\end{table}

\section{Discussion and Limitations}
\textbf{Communication construction versus route selection.} The broad-zoo gap between Core, ABC, and U supports the central thesis that both communication construction and route selection matter. ABC captures exact aggregate-before-communication but not bounded islands, compatible-channel blocking, capacity-aware sink choice, or fused/ordinary route trade-offs. The pass-by-pass study now directly attributes the dominant traffic reduction to the structural stages.

\textbf{Why separate metrics matter.} A compact shared tree can still spend excessive hops or serialize at a bottleneck. Steiner illustrates the failure mode: it has the best synthetic FusedTLU but much worse TLU and latency. Conversely, the last ablation step slightly improves fused footprint while slightly increasing replay latency. Reporting all three metrics exposes these trade-offs instead of burying them in one weighted score.

\textbf{Model boundary.} The deterministic replay excludes router buffers, virtual channels, backpressure, memory hierarchy, RTL timing, area, and energy. Baselines reproduce objective families in a common harness rather than full systems with different hardware assumptions. The model zoo favors breadth of lowerable communication patterns over production-scale datasets and does not include complete masked-softmax attention/GAT lowering. The ablation isolates the global NoC graph after legal local merges; a calibrated implementation must also account for the chosen local-aggregation substrate.

\textbf{Validity and future work.} Frozen instances and deterministic replay avoid run-order noise, but aggregate means do not replace cycle-accurate validation. Future work should calibrate routes in an RTL or cycle-accurate NoC, expose router/buffer parameters, add larger real graph and attention workloads, and extend the formulation to cross-layer residence and one-to-many sharing.

\section{Design Implications}
\textbf{Reduction should precede expensive search.} The pass ablation shows a monotonic decrease in carriers and TLU through the structural stages, while the final route objective produces a much smaller change. A practical compiler should therefore run legal coalescing and blocking before allocating a large route-search budget. Search can then be conditioned on cheap evidence such as a large replay-to-lower-bound gap, concentrated edge pressure, or a candidate-guard disagreement.

\textbf{Locality determines available headroom.} Hotspot and power-law sources expose repeated locations and neighborhoods, so U removes roughly three quarters of their global traffic. Uniform cases still benefit, but less. This suggests exposing locality statistics at lowering time: duplicate-tile rate, island coverage, and compatible-signature frequency can predict whether structural reduction will dominate or whether the mapper should spend effort on placement and congestion avoidance.

\textbf{One weighted score is insufficient.} The Steiner and route-candidate results show two different failure modes. A small fused footprint can coexist with long ordinary routes, while a fused-aware detour can slightly improve FusedTLU and worsen replay latency. Keeping $T_{\mathrm{sched}}$, TLU, and FusedTLU separate lets a deployment choose a policy aligned with its actual substrate rather than inheriting an arbitrary paper weight.

\textbf{The pass should degrade gracefully.} The legality table is not merely defensive documentation. It enables incremental adoption in a production compiler: unsupported operators, full sink tiles, incompatible feature signatures, or a NoC without intermediate arithmetic all fall back to a semantically identical carrier graph. The optimization opportunity grows as hardware capabilities are exposed, but correctness does not depend on enabling every pass.

\section{Conclusion}
MeshReduce-U rewrites legal neural-reduction carrier graphs before routing. Across synthetic and lowerable neural workloads it reduces ordinary link usage and replay latency, while the matched ablation shows that structural rewrites provide the dominant gain.


\begin{thebibliography}{99}

\bibitem{ahuja1993}
Ravindra K. Ahuja, Thomas L. Magnanti, and James B. Orlin.
\newblock \emph{Network Flows: Theory, Algorithms, and Applications}.
\newblock Prentice Hall, 1993.

\bibitem{dally2001}
William J. Dally and Brian Towles.
\newblock Route packets, not wires: On-chip interconnection networks.
\newblock In \emph{Proceedings of the 38th Design Automation Conference}, pages 684--689, 2001.

\bibitem{dijkstra1959}
Edsger W. Dijkstra.
\newblock A note on two problems in connexion with graphs.
\newblock \emph{Numerische Mathematik}, 1:269--271, 1959.

\bibitem{geng2021}
Tong Geng, Chunshu Wu, Yongan Zhang, Cheng Tan, Chenhao Xie, Haoran You, Martin C. Herbordt, Yingyan Lin, and Ang Li.
\newblock I-GCN: A graph convolutional network accelerator with runtime locality enhancement through islandization.
\newblock In \emph{54th Annual IEEE/ACM International Symposium on Microarchitecture}, pages 1051--1063, 2021.

\bibitem{gomory1961}
Ralph E. Gomory and T. C. Hu.
\newblock Multi-terminal network flows.
\newblock \emph{Journal of the Society for Industrial and Applied Mathematics}, 9(4):551--570, 1961.

\bibitem{kou1981}
Lawrence Kou, George Markowsky, and Leonard Berman.
\newblock A fast algorithm for steiner trees.
\newblock \emph{Acta Informatica}, 15:141--145, 1981.

\bibitem{murali2004}
Srinivasan Murali and Giovanni De Micheli.
\newblock Bandwidth-constrained mapping of cores onto NoC architectures.
\newblock In \emph{Design, Automation and Test in Europe Conference and Exhibition}, pages 896--901, 2004.

\bibitem{shivdikar2024}
Kaustubh Shivdikar, Nicolas Bohm Agostini, Malith Jayaweera, Gilbert Jonatan, Jos\'e L. Abell\'an, Ajay Joshi, John Kim, and David R. Kaeli.
\newblock Neurachip: Accelerating GNN computations with a hash-based decoupled spatial accelerator.
\newblock In \emph{51st ACM/IEEE Annual International Symposium on Computer Architecture}, pages 946--960, 2024.

\bibitem{stevens2021}
Jacob R. Stevens, Dipankar Das, Sasikanth Avancha, Bharat Kaul, and Anand Raghunathan.
\newblock GNNerator: A hardware/software framework for accelerating graph neural networks.
\newblock In \emph{58th ACM/IEEE Design Automation Conference}, pages 955--960, 2021.

\bibitem{su2022}
Jiajun Su.
\newblock ABC: Aggregation before communication, a communication reduction framework for distributed graph neural network training and effective partition, 2022.

\bibitem{sun2025}
Guangyu Sun et al.
\newblock Accelerating GNN training through locality-aware dropout and merge, 2025.

\bibitem{tiwari2022}
Binayak Tiwari, Mei Yang, Xiaohang Wang, and Yingtao Jiang.
\newblock Data streaming and traffic gathering in mesh-based NoC for deep neural network acceleration.
\newblock \emph{Journal of Systems Architecture}, 126:102466, 2022.

\bibitem{unnikrishnan2023}
Nanda K. Unnikrishnan, Joe Gould, and Keshab K. Parhi.
\newblock SCV-GNN: Sparse compressed vector-based graph neural network aggregation.
\newblock \emph{IEEE Transactions on Computer-Aided Design of Integrated Circuits and Systems}, 42(12):4803--4816, 2023.

\bibitem{zhang2023dynasparse}
Bingyi Zhang and Viktor K. Prasanna.
\newblock Dynasparse: Accelerating GNN inference through dynamic sparsity exploitation.
\newblock In \emph{IEEE International Parallel and Distributed Processing Symposium}, pages 233--244, 2023.

\bibitem{zhang2023graphagile}
Bingyi Zhang, Hanqing Zeng, and Viktor K. Prasanna.
\newblock Graphagile: An FPGA-based overlay accelerator for low-latency GNN inference.
\newblock \emph{IEEE Transactions on Parallel and Distributed Systems}, 34(9):2580--2597, 2023.

\end{thebibliography}
\end{document}